# Inadequate risk analysis might jeopardize the functional safety of modern systems[1]

By Kaj Hänninen[2], Hans Hansson[3], Henrik Thane[4] and Mehrdad Saadatmand[5]

**Abstract:** In the early 90s, researchers began to focus on security as an important property to address in combination with safety. Over the years, researchers have proposed approaches to harmonize activities within the safety and security disciplines. Despite the academic efforts to identify interdependencies and to propose combined approaches for safety and security, there is still a lack of integration between safety and security practices in the industrial context, as they have separate standards and independent processes often addressed and assessed by different organizational teams and authorities. Specifically, security concerns are generally not covered in any detail in safety standards potentially resulting in successfully safety-certified systems that still are open for security threats from e.g., malicious intents from internal and external personnel and hackers that may jeopardize safety. In recent years security has again received an increasing attention of being an important issue also in safety assurance, as the open interconnected nature of emerging systems makes them susceptible to security threats at a much higher degree than existing more confined products.

This article presents initial ideas on how to extend safety work to include aspects of security during the context establishment and initial risk assessment procedures. The ambition of our proposal is to improve safety and increase efficiency and effectiveness of the safety work within the frames of the current safety standards, i.e., raised security awareness in compliance with the current safety standards. We believe that our proposal is useful to raise the security awareness in industrial contexts, although it is not a complete harmonization of safety and security disciplines, as it merely provides applicable guidance to increase security awareness in a safety context.

## Introduction

Software intensive safety-critical systems have been around for a few decades now and there are well-established approaches for ensuring their safety, essentially originating from safety practices within the aerospace industries. Safety practices in these and other domains are dictated by safety-standards that prescribe how systems should be developed, verified and maintained to minimize risks of accidents during the lifetime of a product. When developing safety related electronic and programmable control systems, there are a number of sector specific standards that need to be considered, for example:

[1] This work was performed in the FIA-PiiA project within the Swedish national Vinnova funded strategic innovation programme Process-industrial IT and Automation; www.piia-sip.se and in the SafeCOP project www.safecop.eu funded from the ECSEL joint undertaking under grant agreement n692529

[2] Mälardalen University (kaj.hanninen@mdh.se)
[3] SICS Swedish ICT Västerås and Mälardalen University (hansh@sics.se)
[4] Safety Integrity AB (henrik.thane@safetyintegrity.se)
[5] SICS Swedish ICT Västerås (mehrdad@sics.se)

- ISO13849 and IEC62061 for machines with moving parts (e.g., Industrial robots),
- ISO26262 for Automotive,
- EN50129/EN50128 for Railway, and
- IEC61508 for generic control systems

These standards outline the requirements and recommendations for the safety work in the respective domain. Many of the sector specific standards that have been developed in recent years stem from the IEC61508 standard. In developing a sector specific standard the IEC61508 has been a fundamental source of inspiration for the developers of sector specific standards. Some concepts from the IEC61508 have been adapted as is, whereas other concepts have been reworked to fit the practices in the specific domains.

**Safety-related systems connected to the Internet**
Traditionally, safety related systems have been closed stand-alone products, but recently they are increasingly interconnected or provided with interfaces to the Internet to allow remote diagnostics or enhanced Infotainment as in the case of modern cars. Allowing external communication is an enabler for many useful and exciting functions and services, but is also potentially dangerous, as it opens up for a whole range of security threats. An example is the remote operation of a Jeep Cherokee[6] via the infotainment interface, allowing remote control of braking and steering. A more classical example is the Stuxnet Worm[7] that specifically targets PLCs, which are used in automation of e.g., machinery on factory assembly lines. Stuxnet is believed to be a jointly built American-Israeli cyber weapon, built to target Iranian centrifuges for separating nuclear material. Further examples include hacked insulin pumps and drug infusion pumps. Hacks of the latter have even prompted warnings from the US FDA[8] resulting in guidance on how to address cyber-security to assure safety of medical devices. There are also examples of hacks causing damage to the environment, including a disgruntled former employee that hacked a water treatment facility in Queensland, Australia, deliberately spilling nearly a million liters of raw sewage into local waterways and parks[9].

Although there are well-established security engineering lifecycles, it should be clear from the above examples that there is a lack of guidance on how to combine and exchange knowledge and results of the work in the disciplines. Moreover, the distinction between safety and security is not always clear. What is clear however is that security threats must be considered in the safety work, as system safety (beyond any reasonable doubt) cannot be established for modern open interconnected systems unless the safety work is extended to take aspects of security into account. Nevertheless, current safety standards do generally not prescribe that security threats shall be evaluated in terms of potential hazards, and consequently there are no requirements on mitigations of those threats.

---

[6] http://www.wired.com/2015/07/hackers-remotely-kill-jeep-highway/
[7] https://en.wikipedia.org/wiki/Stuxnet
[8] http://www.fda.gov/Safety/MedWatch/SafetyInformation/SafetyAlertsforHumanMedicalProducts/ucm456832.htm
[9] http://csrc.nist.gov/groups/SMA/fisma/ics/documents/Maroochy-Water-Services-Case-Study_report.pdf

For instance, the railway standards EN50126, EN50128 and EN50129 describe security as an element that can be considered as a component of RAMS (Reliability, Availability, Maintainability, Safety). However, the consideration of security is outside the scope of the standards (with the exception of security in terms of protection against unauthorized access). The road vehicle standard ISO26262 do not mention security at all. The IEC61508 standard recommends that reasonably foreseeable security threats originating from malevolent or unauthorized action should be analyzed. For guidance on vulnerability and risk analysis the standard refers to IEC62443.

The main reason why security is given less attention in the safety standards is the limit of scope that all functional safety standards prescribe. They are essentially restricted to protection against failures and a degree of foreseeable misuse of the systems, not against intentional abuse and misuse.

**Safety standards are out of synch with recent developments**
Although some safety standards recommend addressing security, it is perfectly possible to get a product approved according to safety standards, while there are still remaining security dependencies that could impact safety and cause an accident. This does not imply that the safety or security works are flawed, contrary it implies that addressing of the interdependencies between them are not regulated or guided enough in normative safety standards. As companies tend to focus on getting an approval by safety-certification authorities, rather than ensuring safety under all conditions including security threats. If this practice is not amended, we are bound to see incidents and accidents in the near future.

In these perspectives the long revision periods of safety standards is a real and serious problem. The committees working on these standards are catching up with new developments in the industry in terms of new technologies, increased complexity, new application domains, etc. As such, the standards are often out-of-date, and have usually a 10-year turnaround time between new versions. Thus, the rapid technical development is in itself a threat to safety. Updating and extending the scope of standards is however not a simple task. Considering that it is non-trivial to understand the interdependencies and differences between modern safety critical and security critical systems, regulating processes and providing recommendations for activities and allocation of responsibilities in assuring risk reduction, becomes extremely difficult.

## Considering security risks in the safety management process

Our claim is that an extended safety approach, considering relevant aspects of security together with safety, is required to amend the current practice. For the approach to be applicable in industrial contexts, and approvable by assessment authorities, the extension must be performed in such way that the proposals complies with the normative safety standards. A harmonization of the activities within the practices has been subject for the research community since the early 90s (see e.g., [2]-[61]; [55] presents an extensive survey of approaches combining safety and security). A problem with extensive harmonization approaches is that they may invalidate recommended normative activities, recommendations or practices in the standards. This implies that they need to be adapted by standardization authorities, i.e., they are generally not directly applicable in an industrial context, unless the normative standards are updated to support them. A typical example of this is the efforts in harmonization of risk classification schemes/procedures of different disciplines.

**Safety and Security in an industrial Case Study**
We have taken initial steps towards development of an extended safety approach in the context of current standards. We considered a wide range of security threats and applied a safety standard (IEC62061 – "Safety of Machinery") and a security standard (ISA/IEC-62443, "Industrial communication networks") on a real industrial case. These standards both have the goal to reduce risk, but have somewhat different approaches to manage risk. We have studied the standards, identified commonalities and differences, and based on the differences we have extended the regular safety risk management to include security threats.

In the following we outline the basic steps to extend the activities of a risk management process that complies with IEC62061.

- **Extending the System Definition:** The system definition is the foundation on which all succeeding functional safety work is based. It defines the scope and intended functionality of the product, its environment, as well as its interfaces. All risk analysis is based on the system definition. Specifically, the hazards identification process starts from the system definition in identifying all hazards that can lead to accidents, incidents, damage or significant financial losses. It is therefore important that the system definition is complete and correct to facilitate the identification of all potential sources of hazards. With the increasing interconnectivity of modern automation control systems, the functional safety system definition must be extended to cover not only failures from within the product itself, but also intentional misuse and sabotage.

    In doing this, the traditional reasoning about sources of hazards (failures and foreseeable misuse) must be extended to also include intentional misuse. This implies that the failure model of the environment and interface parts of the system definition will have to be extended with actors and assets being part of, or interfacing, the system. With the complexity and interconnection of entities within modern systems, the establishment of a system definition is however not trivial. To support the definition process, guidance on typical assets and actors that may affect operations are necessary. Within our work, we identified the following threats and interfaces as important to consider when extending a typical safety system definition (note that the list is not complete in any sense, it is purely for guidance)
    - People (internal and external personnel, design authorities, subcontractors, competitors, litigants, press, hackers, criminals, terrorist etc.)

- Nature and accidents (e.g., fires, storms, floods, transportation accidents etc.)
- Interfaces and assets (e.g. fieldbuses and I/O for system functionality, internal product buses and interfaces, sensors, actuators, configuration interfaces, control interfaces, monitoring interfaces and diagnostics interfaces, maintenance interfaces, testing interfaces and upgrading interfaces, infotainment interfaces, external product interfaces (e.g., authentication and authorization interfaces, session management interfaces, USB interfaces etc.), cellular interfaces and additional assets such as., mobile-enabled devices, printers, USB devices, control centers, cloud services, computers, etc.).

The idea with the guidance list is to identify points allowing tampering with wireless and wired communication links, USB ports, user interfaces, etc. and tampering by personnel (both internal and external) during development, testing, maintenance, production, and operation. These assets and actors must all be included as extensions to the traditional system definition and thus the scope and boundary of a traditional system definition will be extended thereafter.

Establishing a system definition has traditionally been the responsibility of the safety organization. It is clear that this work cannot be performed by the safety organization alone. We cannot expect them to have the overall knowledge of everything included in modern interconnected systems. Instead, establishing a system definition should be a joint work that gathers the organizational teams that together can contribute with their knowledge in an effort to establish a complete and correct definition.

- **Extending the Hazards Analysis**: Hazard analysis is the second most important step (after writing the system definition) in the safety management process of developing safety critical systems. The purpose of the analysis is to identify, quantify, rank, and list hazards that can cause accidents or losses during the lifetime of the product. The hazard analysis can be performed with various techniques, and at different stages of the lifecycle. A preliminary hazard analysis is usually performed in the concept phase before any development has been initiated. The hazard analysis is then refined when more details of the system design emerges, and repeated when performing maintenance. A typical hazard analysis that focuses on safety is guided by experiences from similar projects and different analysis techniques such as fault tree analyses, failure mode and effects analyses, event tree analyses etc. The security domain has similar guidance from e.g., threat models, attack tree analyses etc. In our extension of the hazard analysis, we include security threats in the safety analysis, where we essentially assume that failures are not only stemming from the system itself but also from people with malicious intent. The extended scope of the system definition allows for previously unforeseen safety hazards, and additional ways in which a system might enter a hazardous state, to be identified. This results in a more security-aware safety management process.

- **Risk Classification and Mitigations:** Each hazard that has been identified during the hazard analysis must be classified in terms of risk[10] according to the schemes proposed in the safety standards. The new security related hazards that have been found using the extend hazard analysis have therefore to be classified according to the same scheme. Note that this does not imply that the risk classification proposed by the security standards should be ignored for security risks. The reason to classify the security related safety risks according to a safety scheme serves two main purposes: 1) to assure compliance with the safety standard being used and 2) to assure that all safety risks have been classified according to the same scheme. Note also that risk classification stemming from any reused sources of already identified hazards may have to be re-assessed[11] since the scope of the system definition has changed. All hazards (new as well as old) must then be mitigated with safety measures that are on par with the risk classification of the hazard in order to be able to claim that the risk is tolerable. The functional safety standards mandate different levels of rigor for the development and maintenance process, including techniques and measures to be applied depending on the identified risk level (SIL, ASIL, PL, etc.). A consequence of our extended analysis is that proper mitigations may not be found in the safety standards, but have to be taken from the security standards. Here it is necessary to translate the rigor required between the different domains and standards.

- **Assessing Risk Mitigations:** It is advisable that the safety management process distinguish between hazards discovered from a pure safety perspective, hazards discovered from a security perspective that have safety impact, and hazards discovered from an extended safety perspective that includes security threats. The main reason why the origins of the hazards should be categorized is the fact that this allows risk reduction measures to be more appropriately designed. Where the origins of the hazards are purely safety related (e.g., due to failures, foreseeable misuse etc.) the risk reduction measures, techniques and recommendations in safety standards may be followed. For the other cases the risk reduction process needs to consider if the risk can be reduced according to a safety standard or according to a security standard, or with a combination of both safety and security standards. Note however that in order to be able to certify the product, the development and maintenance process steps required in the safety standard must be followed in all cases when implementing mitigations

---

[10] For example, IEC 62061 mandates a classification procedure that considers the consequence of each hazard, the severity (S) of each hazard, the anticipated rate (F) of occurrence, the anticipated occurrence probability (P) of each hazard and measures for hazard avoidance (A)

[11] This implies that reusing experiences and sources such as preliminary hazard lists from previous projects have to be subject to reassessment and re-classification even if the context and functionality remains the same as for previous analyses

even if the mitigation comes from a security standard.

## Reference system – Case study

To validate our proposed approach we have studied a system for material transportation in a mining environment provided by the PiiA-WROOM[12] project. The system is currently under development and will comprise a set of vehicles that cooperate to transport material in an underground mine. A novel characteristic of the system is that some vehicles can be manually or remotely driven during operation. When remotely driven, an operator in a control room above ground controls the vehicle via wired and wireless networks. An advantage is that when there are still hazardous dust and gases remaining after blasting in the mine, the vehicles can be operated remotely from above ground and begin transporting material before it is safe for humans to be present in the mine. These types of systems, with remote controlled or partially autonomous vehicles, are believed to become a common setup in mines.

We studied the system setup and performed a preliminary hazard analysis according to the IEC62061 safety standard. Typically, these types of systems are analyzed and certified in a modular way, meaning that each vehicle or piece of equipment is analyzed for risks, developed and certified in isolation without considering security threats. To test our method we extended the system definition and performed an extended hazard analysis including security threats from intentional and accidental misuse of the system. We managed, quite easily, to identify new unforeseen safety hazards, and additional ways in which a system might enter a hazardous state. These hazards and the events leading to them would not have been found with a traditional safety approach, i.e., these hazards were not found when using the original scope of the IEC62061 system definition and hazards analysis, since the root causes were not considered to be failures.

The exercise clearly shows that interconnected systems, in this case a system-of-systems, lead to additional hazards when security is considered. The risk classification was done according to the IEC62061 and the security related hazards were also given SIL (safety integrity level) classifications. This allowed us to prioritize the risks and to propose suitable mitigations. Mitigation and countermeasures were chosen from the safety and security domains.

Using this new approach, which in essence is harmonized with IEC62061, allows us to develop, and according to IEC62016, certify a complex interconnected system-of-systems subject to security threats.

## Conclusion/ Summary

We have investigated existing functional safety standards and identified critical shortcomings when they are applied to networked systems: they do not consider security threats that may lead to accidents. Likewise, security standards do not cover safety aspects to the same rigor as the safety standards.

---

[12] http://www.projdb.processitinnovations.se/Aktivitet.aspx?id=219

To remedy this, we propose an approach that considers security threats in the safety work process. The approach can be seen as an initial step towards an integrated approach for safety and security, something that will be needed for keeping risks of accidents and incidents in future networked cooperating products at acceptable levels.

We applied our approach to the IEC62061 functional safety standard in a case study that clearly shows that we systematically can identify additional hazards stemming from security threats that would not have been identified using a traditional safety approach. Since the hazard analysis process is harmonized with the IEC62061 standard we believe that it is now possible to develop and certify complex interconnected system-of-systems according to IEC62016, taking security threats into account. Our approach is general and can be applied also to other standards.